\documentclass[
    ,final            % use final for the camera ready runs
  ]
  {aipproc}

\layoutstyle{6x9}

\newcommand{\cs}{\hat{s}}
\newcommand{\ct}{\hat{t}}
\newcommand{\cu}{\hat{u}}
\newcommand{\beq}{\begin{eqnarray}}
\newcommand{\eeq}{\end{eqnarray}}
\newcommand{\la}{\langle}
\newcommand{\ra}{\rangle}

\begin{document}

\title{Soft-Fermion-Pole Mechanism to Single Spin Asymmetry
in Hadronic Pion Production}

 %13.60.Le for SIDIS

\author{Yuji Koike}{
  address={Department of Physics, Niigata University, Ikarashi, Niigata 950-2181, Japan}
}

\author{Tetsuya Tomita}{
  address={Department of Physics, Niigata University, Ikarashi, Niigata 950-2181, Japan}
}

%\author{<author3>}{
%  address={<common address for author2 and author3>}
%  ,altaddress={<author1 address>} % additional visiting address
%}

\begin{abstract}
Single spin asymmetry (SSA) is a twist-3 observable in the collinear factorization approach.
We present a twist-3 single-spin-dependent cross section formula for the pion
production in $pp$-collision, $p^\uparrow p\to\pi X$, relevant to RHIC experiment.
In particular, we calculate the soft-fermion-pole (SFP) contribution to the
cross section from the quark-gluon correlation functions. 
We show that its effect can be as large as the soft-gluon-pole (SGP) contribution
owing to the large SFP partonic hard cross section, even though the derivative of the SFP
function does not participate in the cross section.

\end{abstract}

\maketitle

%%%%%%%%%%%%%%%%%%%%%%%%%%%%%%%%%%%%%%%%%%%%
%% MAINMATTER
%%%%%%%%%%%%%%%%%%%%%%%%%%%%%%%%%%%%%%%%%%%%

%\section{<A section>}

The observed large single spin asymmetry (SSA) in $p^\uparrow p\to h X$ ($h=\pi, K$ {\it etc})
can not be explained by the usual parton model and perturbative QCD.
In the framework of collinear factorization, SSA is a twist-3 observable and 
can be described in terms of the twist-3 quark-gluon-correlation correlation functions
in the nucleon and/or pion\,\cite{QS91}-\cite{KT072}.  Among various components of the
twist-3 cross sections,
a large contribution is expected to come from the twist-3 distribution functions in
the transversely polarized nucleon. ( See \cite{Koike03}, however.)
There are two independent twist-3 distributions $G_F^a(x_1,x_2)$ and $\widetilde{G}_F^a(x_1,x_2)$
which are defined from a light-cone correlation function %of quark and gluon fields 
in the nucleon $\sim \la \bar{\psi}^aF^{\alpha +}\psi^a\ra$ where 
$\psi^a$ is the quark field of flavor $a$ and $F^{\alpha +}$ is the gluon's field
strength.  For the explicit definition 
of $\{G_F^a(x_1,x_2),\widetilde{G}_F^a(x_1,x_2)\}$ and
the relation between different expressions for the twist-3 distribution functions, see
\cite{EKT06,EKT07}. 
Due to the naively $T$-odd nature of SSA, the relevant cross section occurs
as a pole contribution of an internal propagator in the partonic hard part.
For $p^\uparrow p\to\pi X$, two kinds of poles contribute:
soft-gluon-pole (SGP) leading to $x_1=x_2$ and soft-fermion-pole
(SFP) leading to $x_i=0$ ($i=1$ or 2).  
Thus the contribution to the cross section from these functions can be schematically written as
\beq
  \Delta\sigma^{\rm tw3} &=&    
\biggl(G_F(x,x)-x\frac{dG_F(x,x)}{dx}
    \biggr) \otimes f(x') \otimes D(z) \otimes \hat{\sigma}_{\mathrm{SGP}} \nonumber\\
     &+& G_F(0,x) \otimes f(x') \otimes D(z) \otimes \hat{\sigma}_{\mathrm{SFP}} 
     + \widetilde{G}_F(0,x) \otimes f(x') \otimes D(z) \otimes \hat{\sigma}'_{\mathrm{SFP}},
\label{eq1}
\eeq
where we have used the fact that
the SGP contribution appears in the 
combination of $G_F(x,x)-x{dG_F(x,x)\over dx}$\,\cite{Kouvaris,KT072}.
Note only $G_F(x,x)$ contributes through the SGP, since 
$\widetilde{G}_F(x,x)=0$ due to the anti-symmetric nature 
of $\widetilde{G}_F(x_1,x_2)$ under $x_1\leftrightarrow x_2$. 
The origin of this combination and the connection of
$\hat{\sigma}_{\mathrm{SGP}}$ to the twist-2 cross section was clarified in \cite{KT072,KT071}.
Previous analyses\,\cite{QS99,Kouvaris} focussed on the SGP contribution (first term 
of (\ref{eq1})), assuming that
this one is a dominant contribution.
However, there is no clue that the SFP functions $G_F(0,x)$ and $\widetilde{G}_F(0,x)$
themselves are small, and its importance depends
on the behavior of the partonic hard cross sections $\hat{\sigma}_{\mathrm{SFP}}$
and $\hat{\sigma}'_{\mathrm{SFP}}$. 

The purpose of this report is to present the SFP cross section for
$p^\uparrow p\to\pi X$ and to study its impact 
compared with the SGP contribution, assuming that
the SFP functions $G_F(0,x)$ and $\widetilde{G}_F(0,x)$
have the same order of magnitude as the SGP function $G_F(x,x)$.

For the calculation of the SFP contribution to $p^\uparrow(p,S_\perp)+p(p')\to\pi(P_h)+X$, 
we followed the formalism in \cite{EKT07}.  With the Mandelstam variables
for this process, 
$S=(p+p')^2$, $T=(p-P_h)^2$ and $U=(p'-P_h)^2$, the SFP contribution can be
written as\,\cite{Tomita}:
\beq
 && E_h \frac{d^3\Delta\sigma^{\rm SFP}}{dP_h^3} =\frac{\alpha_s^2}{S}
   \frac{M_N\pi}{2} \ \epsilon^{-+P_h S_{\perp}} 
   \sum_{a,b,c} \int_{z_{min}}^1
   \frac{dz}{z^3}
 \int_{x^{\prime}_{min}}^1\frac{dx^{\prime}}{x^{\prime}}
 \int  \frac{dx}{x} \,  \frac{1}{x^{\prime}S+T/z}\,\nonumber\\
&&\qquad\qquad\times\delta \biggl( x-\frac{-x^{\prime}U/z}
     {x^{\prime}S+T/z} \biggr)
%&&\qquad\quad \times 
\left[
\sum_{a,b,c} \left( G_F^a(0,x) + \widetilde{G}_F^a(0,x) \right) \right.\nonumber\\[5pt]
&&\left.\qquad\times
\left\{ 
q^b(x')\left( D^c(z) \hat{\sigma}_{ab\to c} +   D^{\bar{c}}(z)\hat{\sigma}_{ab\to \bar{c}} 
\right) %\right.\right. \nonumber\\
%&&\left.\left. \hspace{6.5cm} 
+ q^{\bar{b}}(x')
\left( D^c(z)\hat{\sigma}_{a\bar{b}\to c} +   
D^{\bar{c}}(z)\hat{\sigma}_{a\bar{b}\to \bar{c}} \right) 
\right\} 
\right. 
\nonumber \\[5pt]
&&\left. 
\qquad\qquad  + \sum_{a,b} \left( G_F^a(0,x) + \widetilde{G}_F^a(0,x) \right) 
\left(
q^b(x')D^g(z)\hat{\sigma}_{ab\to g} + q^{\bar{b}}(x')D^{g}(z)\hat{\sigma}_{a\bar{b}\to g} 
\right) 
\right. \nonumber\\
&&\left. 
\qquad \qquad + \sum_{a,c} \left( G_F^a(0,x) + \widetilde{G}_F^a(0,x) \right) G(x')\left(
D^c(z)\hat{\sigma}_{ag\to c} + D^{\bar{c}}(z)\hat{\sigma}_{ag\to \bar{c}} 
\right) \right.\nonumber\\
&& 
\qquad\qquad \left. + \sum_a\left( G_F^a(0,x) + \widetilde{G}_F^a(0,x) 
\right)G(x')D^g(z)\hat{\sigma}_{ag\to g}
\right],
\label{SFPfinal}
\eeq
where 
the summation $\sum_{a,b,c}$ implies that the sum of $a$ is over all quark and
anti-quark flavors ($a=u,d,s,\bar{u},\bar{d},\bar{s},\cdots$) and 
the sum of $b$ and $c$ is restricted onto quark flavors when $a$ is a quark and
onto anti-quark flavors when $a$ is an anti-quark.  (For an anti-quark $b$, $\bar{b}$ denotes
quark flavor.)
%x =\frac{-x^{\prime}U/z}{x^{\prime}S+T/z},
The integration region in the convolution formula is specified by
$x^{\prime}_{min} =\frac{-T/z}{S+U/z}$ and $z_{min}= -\frac{(T+U)}{S}$.
$G(x')$ represents the gluon distribution and $D^g(z)$ is the gluon fragmentation function for the pion.
$\hat{\sigma}_{ab\to c}$ {\it etc} represents partonic hard cross sections 
where $c$ is the flavor of the parton fragmenting into the pion.  They
are functions of the 
Mandelstam variables in the parton level; 
$\hat{s}=(xp+x^{\prime}p^{\prime})^2=xx^{\prime}S$, 
$\hat{t}= (xp-P_h/z)^2=\frac{x}{z}T$ and
$\hat{u}=(x^{\prime}p^{\prime}-P_h/z)^2=\frac{x^{\prime}}{z}U$
and are given as follows ($N=3$ is the number of color):
\beq
\hat{\sigma}_{ab\to c} %&=& \hat{\sigma}^7_{(a)}\delta_{ac} +
%\left( \hat{\sigma}^7_{(c)}+\hat{\sigma}^8_{abc} \right) \delta_{ab}\delta_{ac}\nonumber\\
&=&{-(N^2\cs+2\ct)(\cs^2+\cu^2)\over N^2 \ct^3 \cu} \delta_{ac}
+ { -(N^2\ct+\cu-\cs)\cs \over N^3\ct\cu^2}\delta_{ab}\delta_{ac},\nonumber\\
\hat{\sigma}_{ab\to \bar{c}} &=&0,\nonumber\\
\hat{\sigma}_{a\bar{b}\to c} %&=& \hat{\sigma}^7_{(b)}\delta_{ac} 
%+\hat{\sigma}^7_{(d)}\delta_{ab} 
%+\hat{\sigma}^7_{(fg)} \delta_{ab}\delta_{ac} \nonumber\\
&=&{(N^2\cu+2\ct)(\cs^2+\cu^2)\over N^2 \ct^3 \cu} \delta_{ac}
+{(N^2\cu+2\cs)(\ct^2+\cu^2)\over N^2 \cs^2\ct \cu} \delta_{ab}
-{ (N^2-1)\cu^2\over N^3 \cs\ct^2} \delta_{ab}\delta_{ac},\nonumber\\
\hat{\sigma}_{a\bar{b}\to \bar{c}} %&=& \hat{\sigma}^7_{(e)} \delta_{ab}
%+ \left( \hat{\sigma}^7_{(h)}+\hat{\sigma}^8_{(def)}
%\right)\delta_{ab}\delta_{ac}\nonumber\\
&=&{-(N^2\ct+2\cs)(\ct^2+\cu^2)\over N^2 \cs^2\ct \cu} \delta_{ab}
+ { -N^2\cs+\ct-\cu \over N^3\cu^2} \delta_{ab}\delta_{ac},
\eeq
\beq
\hat{\sigma}_{ab\to g} &=& {(N^2\cs+2\ct)(\cs^2+\cu^2)\over N^2 \ct^3 \cu} 
+{-1\over N^3\cs\ct\cu^2}
\left(N^2(\cs^3+3\cs^2\cu-2\cu^3)+\cs^3-\cs^2\cu\right) \delta_{ab},\nonumber\\
\hat{\sigma}_{a\bar{b}\to g} &=& {-(N^2\cu+2\ct)(\cs^2+\cu^2)\over N^2 \ct^3 \cu} \nonumber\\
&&\hspace{-0.8cm}+\left\{{1\over N^3}\left({\cu\over \cs\ct}+{1\over \cu}\right)
+{1\over N}\left( {\cs^2+\cs\ct+\ct^2\over \cs\cu^2}-{\cu\over \ct^2}\right) + 
{ N(\cu^3-\ct^3)(\ct^2+\cu^2)\over \cs^2\ct^2\cu^2}\right\}\delta_{ab},
\eeq
\beq
\hat{\sigma}_{ag\to c} &=& 
\left\{
{ N^2(\cs^3-\cu^3)(\cs^2+\cu^2) \over (N^2-1)\cs\ct^3\cu^2}\right.\nonumber\\
&&\left.\hspace{-1.2cm}
+{ \cs\cu(\cs^2+\cs\cu-\cu^2) - N^2(\cs^4+\cs^3\cu+\cs^2\cu^2+\cs\cu^3+\cu^4)
\over N^2(N^2-1)\cs\ct^2\cu^2} \right\}
\delta_{ac}%\nonumber\\
%&& 
+{ (N^2\cu+2\cs)(\ct^2+\cu^2) \over N(N^2-1)\cs^2\ct\cu},\nonumber\\
\hat{\sigma}_{ag\to \bar{c}} &=&
 {\cs+2\ct-N^2\cs \over N^2(N^2-1)\cu^2} \delta_{ac} +
{ -(N^2\ct+2\cs)(\ct^2+\cu^2) \over N(N^2-1)\cs^2\ct\cu},
\eeq
\beq
\hat{\sigma}_{ag\to g} &=&
{ -N^2\over (N^2-1)\cs^2\ct^3\cu^2}
\left( 4\cs^6+11\cs^5\ct +19\cs^4\ct^2+22\cs^3\ct^3+19\cs^2\ct^4+11\cs\ct^5+4\ct^6
\right) \nonumber\\
&&+ { 1\over N^2(N^2-1)\cs\ct^2\cu^2}
\left\{ -\cs\ct\cu^2 +N^2(\cs^4+\cs^3\ct+2\cs^2\ct^2+\cs\ct^3+\ct^4)\right\}.
\eeq
A remarkable feature of (\ref{SFPfinal}) is that the partonic hard cross sections 
for $G_F(0,x)$ and $\widetilde{G}_F(0,x)$ are the same, even though
each diagram gives different contributions 
for the two functions\,\cite{Tomita}.  Accordingly, they appear in
the combination of $G_F^a(0,x) +\widetilde{G}_F^a(0,x)$ in (\ref{SFPfinal}).
We also note that some terms in the hard cross section (the first terms in 
$\hat{\sigma}_{ab\to c}$,
$\hat{\sigma}_{ab\to g}$, $\hat{\sigma}_{ag\to c}$, 
$\hat{\sigma}_{ag\to g}$ etc) accompany the large color factors compared
to the SGP cross section derived in \cite{Kouvaris}, and show the steep rising behavior in the
forward direction (i.e. large $S/T$).  This suggests that the SFP contribution
gives rise to the nonnegligible contribution to the asymmetry.

To see the impact of the SFP contribution, we have performed 
a numerical calculation of $A_N$ for $p^\uparrow p\to\pi X$, assuming
that the SFP function is of the same order of magnitude as the SGP function.
Kouvaris {\it et al.} \cite{Kouvaris} have parametrized the
SGP function $G_F^a(x,x)$ so that the SGP contribution reproduces
the $A_N$ data obtained from RHIC and FNAL data.  Their analysis shows 
that both data are reasonably well reproduced. In particular, they found that 
the derivative term in (\ref{eq1})
brings dominant contribution compared to the nonderivative contribution. 
Here we adopt Fit(I) of \cite{Kouvaris} and assume $G_F^a(0,x)+\widetilde{G}_F^a(0,x)=G_F^a(x,x)$
($a=u,d$).  
Fig. 1 shows $A_N$ for the pion at $\sqrt{S}=200$ GeV and $P_{hT}=1.5$ GeV
with and without SFP contribution.  As is seen from the figure that
the SFP contribution brings large effect in the positive $x_F$ region,
while its effect is negligible in the negative $x_F$ region.
From this figure it is clear that
the SFP contribution can affect $A_N$ significantly 
even though it does not receive enhancement by the derivative unlike SGP contribution,
unless the SFP function itself is small.  

To summarize, we have calculated the SFP contribution to
the cross section for $p^\uparrow p\to h X$ 
associated with the twist-3 quark-gluon correlation functions in the polarized nucleon, 
and have shown that its effect is significant and should be included in the analysis
of $A_N$.  
For a more complete analysis, one needs to include the contribution from
the triple-gluon twist-3 distribution function as well.  Recent study\,\cite{Kang:2008ih}
shows that $A_N$ for the open charm production 
has a potential to determine the function.  We hope that
the global analysis of all those data including all the QCD effects 
will clarify the origin of observed SSA.

%%%%%%%%%%%%%%%%%%%%%%%%%%%%%%%%%%%%%%%%%%%%
%% Sample figure:
%%
%% The option [height=...] scales the picture to the given height,
%% without it it would be printed at its nominal size
%%%%%%%%%%%%%%%%%%%%%%%%%%%%%%%%%%%%%%%%%%%%

\begin{figure}
  \includegraphics[height=.35\textheight]{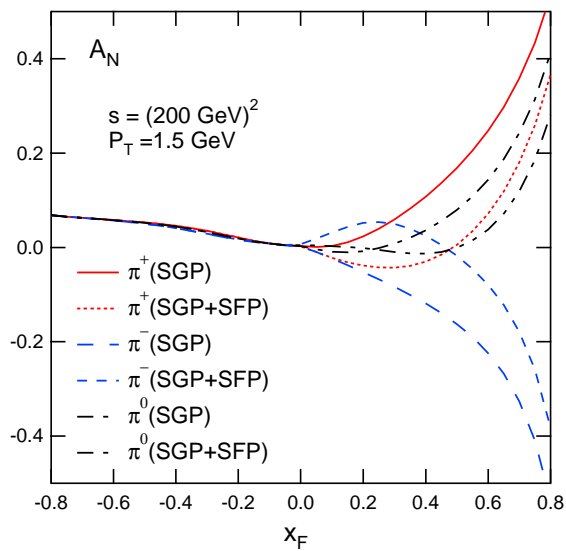}
\caption{$A_N$ for $p^\uparrow p\to\pi X$ at $\sqrt{S}=200$ GeV and $P_{hT}=1.5$ GeV.
Solid, long-dashed, and dash-dot lines are, respectively, $A_N$ for $\pi^+$, $\pi^-$ and $\pi^0$
obtained with only the SGP contribution.  
Dotted, short-dashed, and dash-double-dot lines are, respectively, 
$A_N$ for $\pi^+$, $\pi^-$ and $\pi^0$
obtained with both SGP and SFP contributions.  }
\end{figure}

%%%%%%%%%%%%%%%%%%%%%%%%%%%%%%%%%%%%%%%%%%%%%%%%
%% BACKMATTER
%%%%%%%%%%%%%%%%%%%%%%%%%%%%%%%%%%%%%%%%%%%%%%%%

\begin{theacknowledgments}
The work of Y.K. is supported in part by Uchida Energy Science Promotion Foundation.
\end{theacknowledgments}




%\bibliography{sample}

\IfFileExists{\jobname.bbl}{}
 {\typeout{}
  \typeout{******************************************}
  \typeout{** Please run "bibtex \jobname" to optain}
  \typeout{** the bibliography and then re-run LaTeX}
  \typeout{** twice to fix the references!}
  \typeout{******************************************}
  \typeout{}
 }

\end{document}